\documentclass[preprint,12pt]{elsarticle}
\usepackage{amssymb}
\usepackage{amsthm}
\usepackage{framed} % To put a box to the nomenclature
\usepackage{amsmath,color}
\usepackage{mathrsfs}
\usepackage{graphicx}
\usepackage{epstopdf}
\usepackage{float}
\usepackage{caption}
\usepackage{subcaption}
\usepackage{bm}
\usepackage{bbm}
\usepackage{mathrsfs}
\usepackage{cleveref}
\usepackage{soul}
\usepackage{accents}
\usepackage{color,soul} %to highlight text
\usepackage{color} %To highlight references in the text
\usepackage{nomencl} %to add the nomenclature
\makenomenclature %to add the nomenclature
\biboptions{sort&compress}
\soulregister\citep7 % To allow citep to be inside of highlighted text
\soulregister\citet7 % Idem with citet
\soulregister\citealp7 % Idem with citealp
\newsavebox{\measurebox} %To create a 1 column figure with a 2 column figure on its side
\usepackage{titlesec} %To use paragraph as subsubsubsection
\usepackage{tabu} % To increase the vertical space between cells in tables
\usepackage{longtable}
\usepackage[euler]{textgreek} % To type Greek letters outside of math mode and employing a different font

\journal{Materials}

\makeatletter
\def\@author#1{\g@addto@macro\elsauthors{\normalsize%
    \def\baselinestretch{1}%
    \upshape\authorsep#1\unskip\textsuperscript{%
      \ifx\@fnmark\@empty\else\unskip\sep\@fnmark\let\sep=,\fi
      \ifx\@corref\@empty\else\unskip\sep\@corref\let\sep=,\fi
      }%
    \def\authorsep{\unskip,\space}%
    \global\let\@fnmark\@empty
    \global\let\@corref\@empty  %% Added
    \global\let\sep\@empty}%
    \@eadauthor={#1}
}
\makeatother

\titleformat{\paragraph}
{\normalfont\normalsize\itshape}{\theparagraph}{1em}{}
\titlespacing*{\paragraph}
{0pt}{3.25ex plus 1ex minus .2ex}{1.5ex plus .2ex}

\begin{document}

\begin{frontmatter}



\title{On the Finite Element Implementation of Functionally Graded Materials}


\author{Emilio Mart\'{\i}nez-Pa\~neda\corref{cor1}\fnref{Cam}}
\ead{mail@empaneda.com}

\address[Cam]{Department of Engineering, University of Cambridge, CB2 1PZ Cambridge, UK}

\cortext[cor1]{Corresponding author.}

\begin{abstract}
We investigate the numerical implementation of functionally graded properties in the context of the finite element method. The macroscopic variation of elastic properties inherent to functionally graded materials (FGMs) is introduced at the element level by means of the two most commonly used schemes: (i) nodal based gradation, often via an auxiliary (non-physical) temperature-dependence, and (ii) Gauss integration point based gradation. These formulations are extensively compared by solving a number of paradigmatic boundary value problems for which analytical solutions can be obtained. The nature of the notable differences revealed by the results is investigated in detail. We provide a user subroutine for the finite element package ABAQUS to overcome the limitations of the most popular approach for implementing FGMs in commercial software. The use of reliable, element-based formulations to define the material property variation could be key in fracture assessment of FGMs and other non-homogeneous materials.
\end{abstract}


\end{frontmatter}



\section{Introduction}
\label{Sec:Introduction}

There is an emerging interest in the analysis of the mechanical response of materials with spatially varying properties. New manufacturing technologies make it possible to engineer materials with functionally graded microstructures, so-called functionally graded materials (FGMs). The resulting spatial variation of material properties eliminates stress discontinuities at material interfaces and optimizes material performance under non-uniform service conditions. For example, the performance of coatings subjected to large thermal gradients can be significantly improved by using metal-ceramic FGMs \cite{Kawasaki1987}, which combine the thermal and corrosive resistance of ceramics with the mechanical strength and high tenacity of metals. In addition, FGMs are now employed in a host of commercial applications, ranging from cutting tools to biomedical devices \cite{Uemura2003}. This widespread use of FGMs is largely due to their capacity to reduce residual stresses \cite{Lee1994}, increase the strength of joints \cite{Ramaswamy1997}, and tailor material microstructure to specific service requirements \cite{Marur1998}.

The complexity and cost associated with the manufacture and testing of functionally graded specimens has intensified the use of numerical tools to analyse their mechanical response. Although a variety of numerical techniques have been used, including mesh-free methods \cite{Rao2003,Liu2013a} and enriched formulations \cite{Comi2007,Natarajan2011}, the finite element method is by far the most popular approach \cite{Bao1995,Rousseau2000,Santare2000,Kim2002,Valizadeh2013,IJMMD2015}. Several formulations have been proposed to accurately capture a smooth material gradient by defining the material property variation at the element level \cite{Rousseau2000,Santare2000,Kim2002}. While these formulations are been widely and indistinctly used, a performance assessment of the different types of \emph{graded} elements has not been conducted yet. We investigate the performance of different types of functionally \emph{graded} elements by comparing with analytical solutions of paradigmatic boundary value problems. We show that notable differences can be attained and that the most common approach in commercial software entails a number of limitations. An alternative implementation is presented in the context of the commercial finite element package ABAQUS. 

%%%%%%%%%%%%%%%%%%%%%%%%%%%%%%%%%%%%%%%%%%
\section{Numerical Formulation}
\label{Sec:NumFormulation}

The assignment of material properties in the numerical model must reflect the material property distribution in the functionally graded specimen under consideration. However, an accurate characterization of the material gradient is not a straightforward task. Typically, the information available is the spatial variation of the volume fractions of constituent materials, which is provided as input to the production technique \cite{Butcher1998}. The macroscopic material property variation does not tend to mirror the volume fraction profile, but one can estimate the former from the latter by using homogenization laws \cite{CS2018}. However, the micromechanical assumptions upon which these theoretical mixing laws are built may hinder an accurate characterization of the macroscopic variation of material properties. An alternative approach is to determine the material property variation directly by experimentation. For example, by producing and testing individual homogeneous specimens with a range of volume fractions \cite{Carrillo-Heian2001}, by testing the graded material through indentation or ultrasonic techniques \cite{Krumova2001}, or by cutting and testing small samples from a larger graded specimen \cite{Abanto-Bueno2006}. Capturing this material gradation profile in the numerical model is key to designing optimal FGM specimens, as~well as reproducing and gaining insight into experimental results. 

\begin{figure}[H]
\centering
\includegraphics[scale=1]{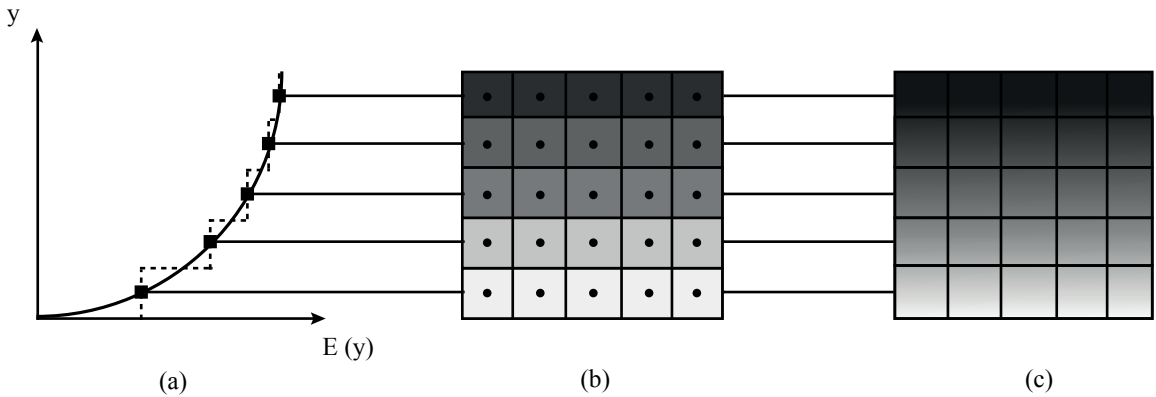}
\caption{Sketch outlining the (\textbf{a}) gradual variation of Young's modulus $E$ along the $y$-coordinate, as captured by (\textbf{b}) \emph{homogeneous} elements and (\textbf{c}) \emph{graded} elements.}
\label{fig:SketchGradHom}
\end{figure}

From the numerical perspective, material properties can vary between elements or between nodes and integration points. Numerical works in the mechanics of functionally graded materials can be classified into two large groups depending on their approach to the implementation of the material gradient, see Figure \ref{fig:SketchGradHom}, using either \emph{homogeneous} elements (see, e.g., \cite{Bao1995}) or \emph{graded} elements (see \cite{IJMMD2015} and references therein). The former is appropriate for layered functionally graded composites, but it constitutes a poor approximation otherwise. Assuming constant material properties within each element leads to a discontinuous step-type variation and requires uniform meshing along the material gradation direction. A material property variation at the element level is generally more appropriate and different graded elements formulations have been proposed. 

\subsection{Gauss Integration Point-Based Variation}
\label{Sec:GaussBased}

Consider an isoparametric finite element with $n$ number of nodes, the displacement field $\bm{u} (\bm{x}) $ is interpolated from the nodal values $\hat{\bm{u}}_i$ as 
\begin{equation}
\bm{u} = \sum_{i=1}^n N_i (\xi, \, \eta, \, \zeta) \, \hat{\bm{u}}_i,
\end{equation}

\noindent where $i$ is a given node and $N_i$ are the shape functions. For example, in an eight-node quadrilateral element, the shape functions read, for the corner nodes
\begin{equation}
N_i = \frac{1}{4} \left( 1 + \xi \xi_i \right) \left( 1 + \eta \eta_i \right) \left( \xi \xi_i + \eta \eta_i -1 \right)
\end{equation}

\noindent and for the mid-side nodes,
\begin{equation}
N_i = \frac{1}{2} \left( 1 - \xi^2 \right) \left( 1+\eta \eta_i \right), \,\,\,\,\,\,\,\,\,\,\,\,\,\,\,\,\,\,  \textnormal{for} \,\,\, \xi_i=0,
\end{equation}
\begin{equation}
N_i = \frac{1}{2} \left( 1 + \xi \xi_i \right) \left( 1 - \eta^2 \right), \,\,\,\,\,\,\,\,\,\,\,\,\,\,\,\,\,\,  \textnormal{for} \,\,\, \eta_i=0,
\end{equation}

\noindent with $\left(\xi , \, \eta \right)$ denoting the intrinsic coordinates in the interval $[-1, \, 1]$, and $\left(\xi_i , \, \eta_i \right)$ denoting the local coordinates of node $i$. Accordingly, the strain field $\bm{\varepsilon} (\bm{x}) $ is computed from the nodal displacements by means of the strain-displacement matrix
\begin{equation}
\bm{\varepsilon} = \sum_{i=1}^n \bm{B}_i \left( \xi, \, \eta, \, \zeta \right) \, \hat{\bm{u}}_i
\end{equation}

\noindent with the matrix $\bm{B}_i$ containing the appropriate derivatives of the shape functions $N_i$. For example, in a plane strain element, the strain-displacement matrix for a node $i$ reads
\begin{equation}
\bm{B}_i \left( \xi, \, \eta \right) = \begin{bmatrix}
  \partial N_i \left( \xi, \, \eta \right) / \partial x & 0 \\
  0 & \partial N_i \left( \xi, \, \eta \right) / \partial y \\
  0 & 0 \\
  \partial N_i \left( \xi, \, \eta \right) / \partial y & \partial N_i \left( \xi, \, \eta \right) / \partial x 
 \end{bmatrix}
\end{equation}

\noindent so as to compute the strain components $\varepsilon_{xx}, \varepsilon_{yy}, \varepsilon_{zz}, \gamma_{xy}$ from the nodal displacements.

Let us assume linear elastic behaviour, which is arguably appropriate for ceramic-based FGMs. The Cauchy stress field $\bm{\sigma}$ is related to the strain tensor $\bm{\varepsilon}$ through a \emph{spatially varying} constitutive matrix $\bm{C} \left(\bm{x} \right)$ as
\begin{equation}\label{Eq:HookeLaw}
\bm{\sigma} = \bm{C} \left(\bm{x} \right) \bm{\varepsilon}.
\end{equation}

The principle of virtual work yields the relation between the deformation work given by the element and the elemental nodal force vector $\bm{F}^e$ as
\begin{equation}\label{Eq:ElSystem}
\bm{K}^e \hat{\bm{u}}_i = \bm{F}^e,
\end{equation}

\noindent where $\bm{K}^e$ is the element strain-displacement matrix---see, for example, Ref. \cite{Bower2009a}. Accordingly, the element stiffness matrix over the volume of the element $V^e$ reads 
\begin{equation}\label{Eq:Ke}
\bm{K}^e = \int_{V_e} {\bm{B}^e}^T \bm{C} \left(\bm{x} \right) \bm{B}^e \, \text{d} V,
\end{equation}

\noindent where $\bm{B}^e$ is the element stiffness matrix, given by the assembly of $\bm{B}_i$ over $n$ nodes. Therefore, the~linear elastic stiffness matrix is defined to match the material gradation profile at the Gauss integration points. This basic finite element formulation for functionally graded solids was presented by Santare and Lambros \cite{Santare2000}.

\subsection{Nodal-Based Variation via Temperature Dependence}
\label{Sec:Nodebased}

An alternative approach to develop a formulation for graded finite elements was proposed by Kim and Paulino \cite{Kim2002}. They propose a generalized isoparametric finite element formulation where the same shape functions are employed to interpolate the nodal displacements, the geometry, and the material properties. Thus, consider a standard isoparametric formulation where the spatial coordinates $(x, y, z)$ are interpolated as
\begin{equation}
x= \sum_{i=1}^n N_i x_i, \,\,\,\,\,\, y= \sum_{i=1}^n N_i y_i, \,\,\,\,\,\, z= \sum_{i=1}^n N_i z_i.
\end{equation}

The isoparametric concept can be generalized to interpolate the spatially varying Young's modulus $E \left( \bm{x} \right)$ and Poisson's ratio $\nu \left( \bm{x} \right)$ as
\begin{equation}
E = \sum_{i=1}^n N_i E_i, \,\,\,\,\,\,\,\,\,\,\,\, \nu = \sum_{i=1}^n N_i \nu_i,
\end{equation}
 
\noindent where $E_i$ and $\nu_i$ are the elastic properties defined at each node $i$. Hence, the material gradient is defined precisely at the nodes and subsequently interpolated to the Gauss integration points to compute the stresses through Equation (\ref{Eq:HookeLaw}). 

\begin{figure}[H]
\centering
\includegraphics[scale=1]{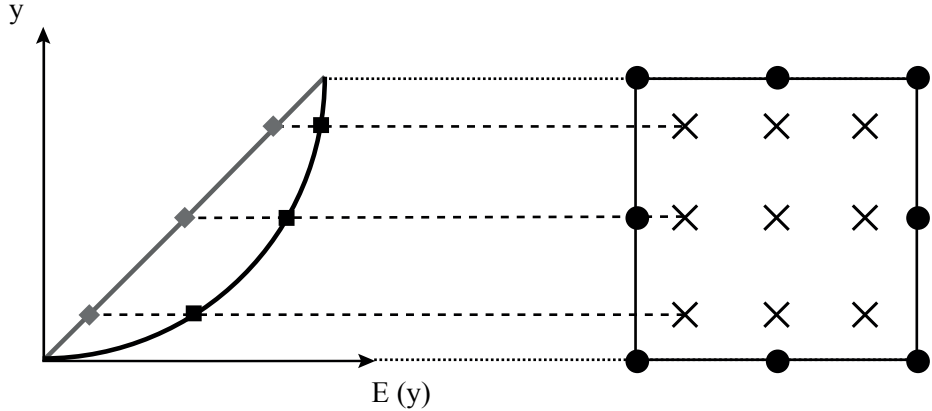}
\caption{Sketch outlining the gradual variation of Young's modulus $E$ and its associated interpolation by means of temperature-based generalized isoparametric graded element for an equivalent interpolation of thermal and mechanical strains.}
\label{fig:SketchLinearApprox}
\end{figure}

A generalized isoparametric graded finite element can be easily implemented into a commercial finite element package by taking advantage of the possibility of defining temperature-dependent material properties \cite{Rousseau2000,IJMMD2015}. For example, one can define $E$ as a function of the temperature and provide the specimen with an initial temperature distribution that matches the Young's modulus variation desired. Here, the temperature has no physical meaning and unwanted thermal strains are suppressed by assigning a zero thermal expansion coefficient. Since the temperature field is defined at the nodes and subsequently interpolated to the Gauss integration points, this technique constitutes a straightforward implementation of a generalized isoparametric graded element, enjoying great popularity. Evident drawbacks are the inability to (i) model thermomechanical problems, and (ii) define different profiles for Poisson's ratio and Young's modulus. Furthermore, to obtain a consistent variation of mechanical and thermal strains, many commercial codes interpolate nodal temperature values using shape functions one order lower than those used for the nodal displacements. Consequently, there is an inherent error in the presence of a nonlinear material gradation profile, as~sketched in Figure \ref{fig:SketchLinearApprox}. The implications of adopting this technique, relative to the Gauss-based approach defined in Section \ref{Sec:GaussBased}, are explored here.

%%%%%%%%%%%%%%%%%%%%%%%%%%%%%%%%%%%%%%%%%%
\section{Results}

\begin{figure}[H]
\centering
\includegraphics[scale=0.8]{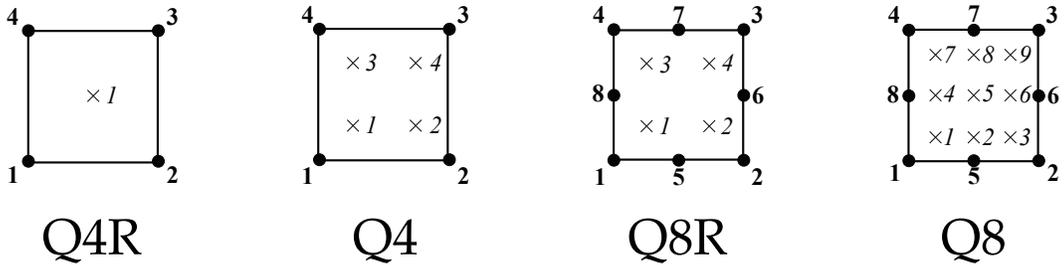}
\caption{Graded finite elements employed and notation used.}
\label{fig:ElementTypes}
\end{figure}

The variation in elastic properties inherent to FGMs is implemented at the element level by making use of user subroutines within the commercial finite element package ABAQUS. The graded elements described in Section \ref{Sec:NumFormulation} can be readily implemented by using a USDFLD subroutine, for a Gauss points-based implementation, or a UFIELD subroutine, for a nodal-based graded element. In~addition, as discussed above, the temperature can be used as an auxiliary field to effectively implement a generalized isoparametric graded element. The user must provide the material properties as a function of a user defined field (or temperature). Then, a suitable field (or temperature distribution) is defined to match the material property variation desired. A direct comparison between the two approaches in terms of computational time is hindered by their different implementations; nevertheless, the sampling of material properties at integration points or nodes is achieved at a negligible computational cost. The performance of different types of graded elements will be benchmarked by considering a Gauss point-based implementation (Section \ref{Sec:GaussBased}) and, via temperature dependent properties, a generalized isoparametric  approach (Section \ref{Sec:Nodebased}). For the sake of simplicity, we will consider bi-dimensional problems and four quadrilateral element types (see Figure \ref{fig:ElementTypes}): linear elements with reduced integration (Q4R), linear elements with full integration (Q4), quadratic elements with reduced integration (Q8R), and quadratic elements with full integration (Q8).  

\begin{figure}[H]
        \centering
        \begin{subfigure}[h]{0.22\textwidth}
                \centering
                \includegraphics{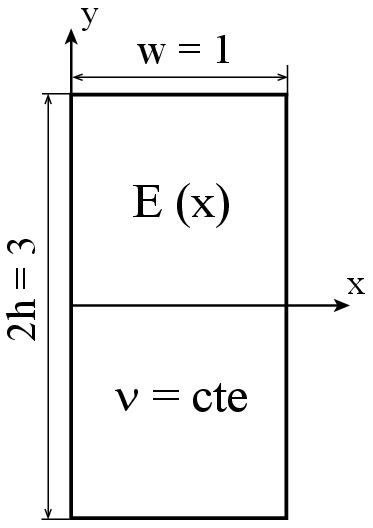}
                \caption{}
                \label{fig:Fig3a}
        \end{subfigure}
        \begin{subfigure}[h]{0.22\textwidth}
                \centering
                \includegraphics{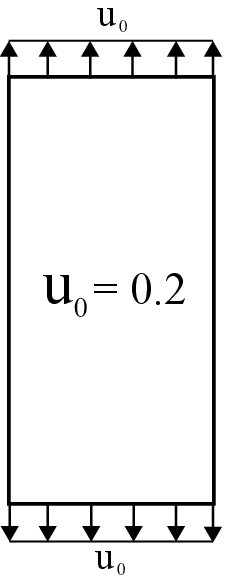}
                \caption{}
                \label{fig:Fig3b}
        \end{subfigure}
        \begin{subfigure}[h]{0.22\textwidth}
                \centering
                \includegraphics{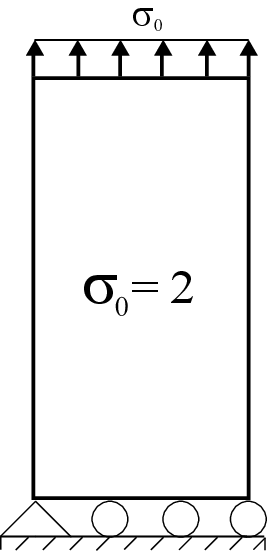}
                \caption{}
                \label{fig:Fig3c}
        \end{subfigure}
        \begin{subfigure}[h]{0.22\textwidth}
                \centering
                \includegraphics{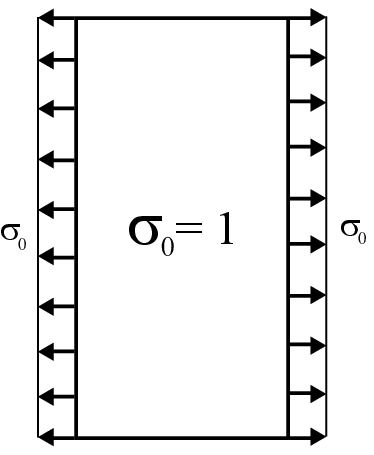}
                \caption{}
                \label{fig:Fig3d}
        \end{subfigure}
        \vspace{+3pt}

        \caption{Boundary value problems under consideration: (\textbf{a}) functionally graded plate with spatially varying Young's modulus subjected to (\textbf{b}) uniform displacement perpendicular to the material gradient direction, (\textbf{c}) uniform traction perpendicular to the material gradient direction, and (\textbf{d}) uniform traction in the direction parallel to material gradation---consistent units.}\label{fig:BoundaryValueProblems}
\end{figure}

Three plane problems for which analytical solutions can be obtained will be addressed, as shown in Figure \ref{fig:BoundaryValueProblems}. Young's modulus will be varied along the $x$-direction and Poisson's ratio will be assumed to be constant. We assume plane stress conditions. As sketched in Figure \ref{fig:BoundaryValueProblems}, the three boundary value problems considered involve a functionally graded plate being subjected to (i) uniform displacement perpendicular to the material gradient direction, (ii) uniform traction perpendicular to the material gradient direction, and (iii) uniform traction in the direction parallel to material gradation.

In all cases, we assume that the Young's modulus varies exponentially as
\begin{equation}\label{eq:Evariation}
E(x)=E_0 \exp \left( \beta x \right)
\end{equation}

\noindent with $E_0$ and $\beta$ being material constants. \textcolor{black}{This choice is motivated by the existence of analytical solutions for functionally graded solids exhibiting an exponential variation of the elastic properties; see, for~example, Refs. \cite{Erdogan1997,Kim2002}. Many other functions have been employed in the literature (see Ref.~\cite{Sayyad2019} for a review), but our choice is appropriate for our aim: comparing on equal footing different graded finite element implementations}. Consistent units are throughout the manuscript and, therefore, units will be omitted subsequently. We consider a width $w=1$, a total height of $2h=3$, and choose $E_0=1$ and $\beta=\ln 8$ so as to vary $E$ gradually in the $x$-direction from $E(0)=1$ to $E(w)=8$.

\subsection{Uniform Displacement Perpendicular to the Material Gradient Direction}
\label{Sec:DispPerpendicular}

Consider first the case of an FGM plate subjected to a remote strain $\varepsilon_0=u_0/h$, where $u_0$ denotes the displacement in the remote boundary and $h$ denotes half the height of the plate. A constant Poisson's ratio of $\nu=0.3$ throughout the plate is assumed. The relevant stress component is given by
\begin{equation}
\sigma_{yy} \left(x, y \right) = E(x) u_0 / h.
\end{equation}

\begin{figure}[H]
\makebox[\linewidth][c]{%
        \begin{subfigure}[b]{0.45\textwidth}
                \centering
                \includegraphics[scale=0.8]{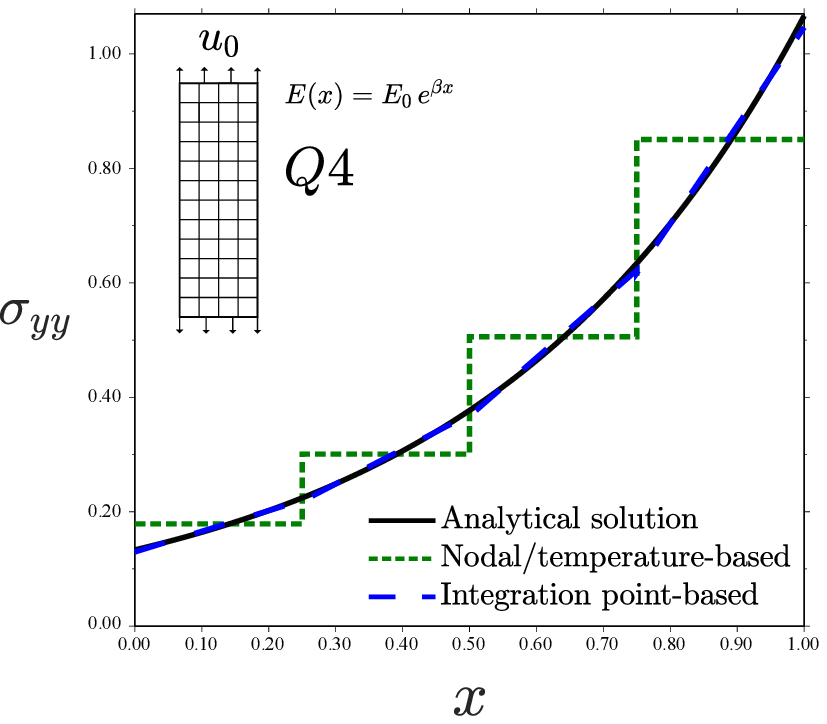}
                \caption{}
                \label{fig:Case1Q4}
        \end{subfigure}
        \begin{subfigure}[b]{0.45\textwidth}
                \raggedleft
                \includegraphics[scale=0.8]{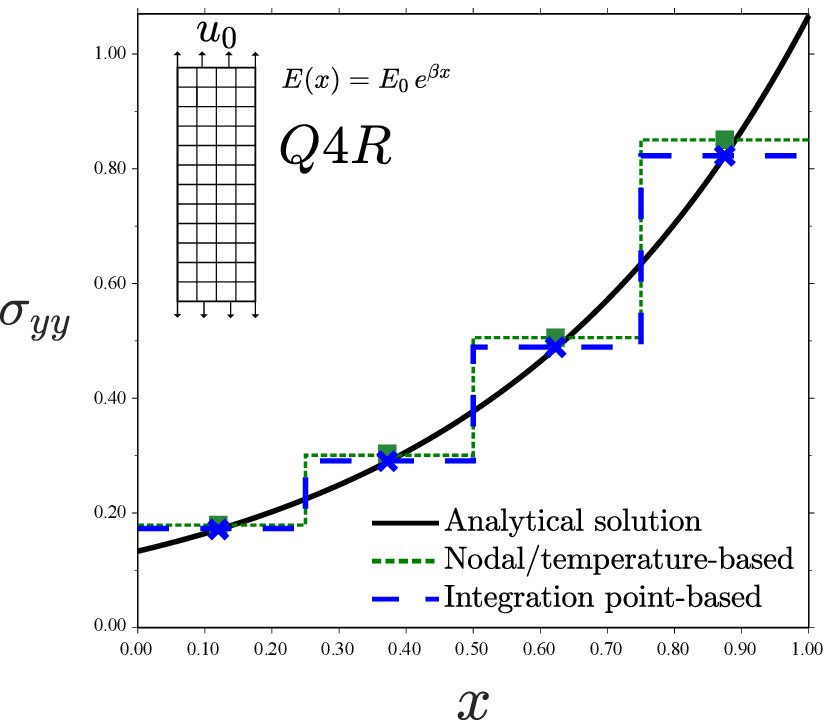}
                \caption{}
                \label{fig:Case1Q4R}
        \end{subfigure}}\\

\makebox[\linewidth][c]{%       
        \begin{subfigure}[b]{0.45\textwidth}
                \centering
                \includegraphics[scale=0.8]{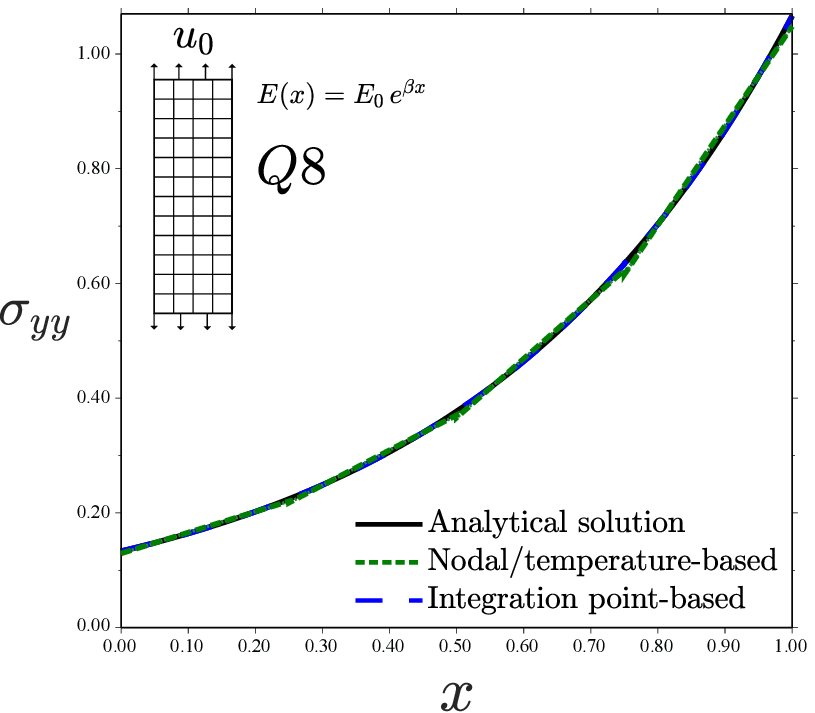}
                \caption{}
                \label{fig:Case1Q8}
        \end{subfigure}
        \begin{subfigure}[b]{0.45\textwidth}
                \raggedleft
                \includegraphics[scale=0.8]{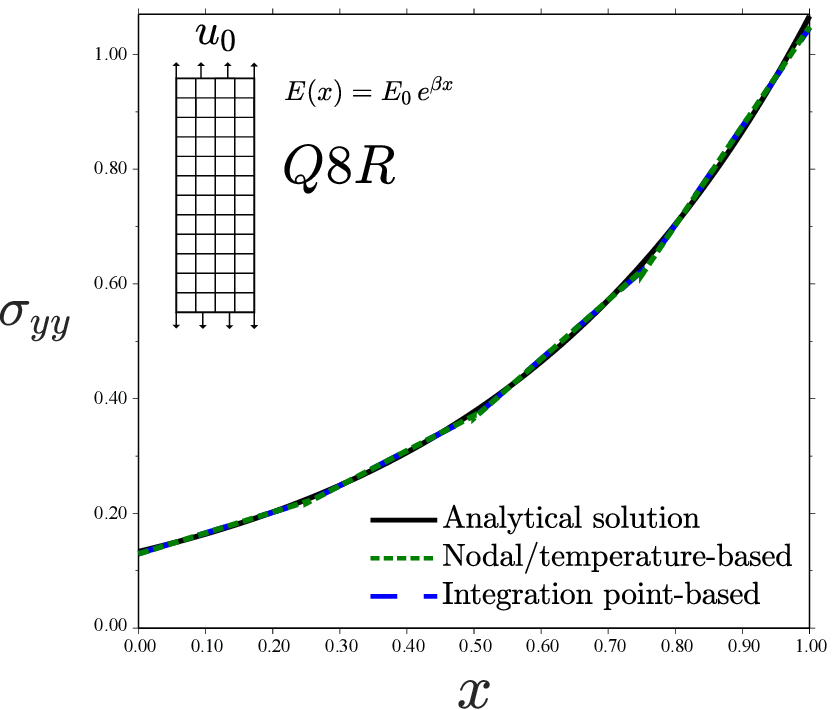}
                \caption{}
                \label{fig:Case1Q8R}
        \end{subfigure}
        }       
        \caption{Uniform displacement perpendicular to the material gradient for different kinds of elements: (\textbf{a}) Q4; (\textbf{b}) Q4R; (\textbf{c}) Q8; and (\textbf{d}) Q8R---consistent units.}\label{fig:Case1}
\end{figure}

This analytical solution is compared in Figure \ref{fig:Case1} with the finite element results obtained for the element types and graded element formulations described above. A remote displacement of $u_0=0.2$ is prescribed in all cases. As shown in the insets of Figure \ref{fig:Case1}, a uniform mesh of 4 by 12 elements is employed.

\begin{figure}[H]
\centering
\includegraphics[scale=1.4]{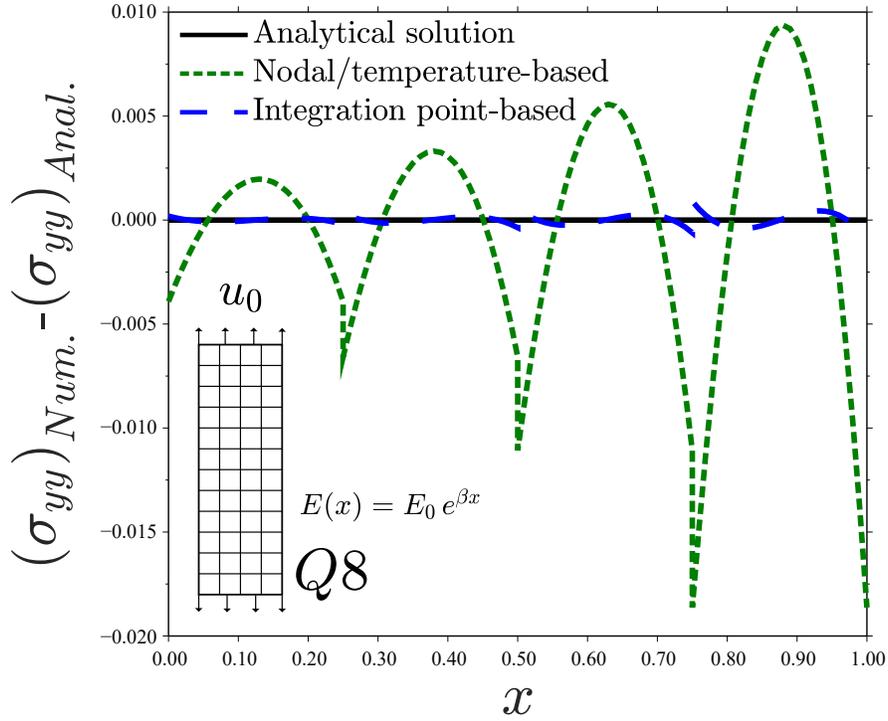}
\caption{Uniform displacement perpendicular to the material gradient. Error analysis for the Q8 element. Consistent units.}
\label{fig:Case1Q8error}
\end{figure}

Results reveal differences between the different types of graded elements. Consider first the case of the linear element with full integration Q4---see Figure \ref{fig:Case1}a. The Gauss integration point-based approach accurately captures the material gradient depicted by the analytical solution. In fact, the~numerical result is exact at the integration points since the displacement field is linear; a single Q4 element will suffice to capture the FGM response. However, the generalized isoparametric formulation (via temperature-dependent properties) exhibits a step-type variation with constant stress in each element. This behaviour is inherently related to how ABAQUS interpolates nodal temperature values. As many other finite element packages, ABAQUS interpolates nodal temperatures with shape functions that are one order lower than those used for the displacements, so as to obtain an equivalent distribution of mechanical and thermal strains. An average value of the temperature in the nodes is passed to the integration points when using linear elements and a linear variation is assumed in quadratic elements. In agreement with expectations, the use of linear elements with reduced integration (Q4R, see Figure~\ref{fig:Case1}b) exhibits the response inherent to homogeneous elements for both cases. However, the constant value of $\sigma_{yy}$ attained in each element depends on the implementation approach. The~integration point-based scheme computes the exact $\sigma_{yy}$ at the element centroid, where $E$ is sampled. On the other hand, the nodal-based approach averages nodal temperatures, introducing a source of error when the elastic properties vary in a nonlinear manner. The use of quadratic elements leads to a good agreement with the analytical solution for both schemes, although differences are observed (Figure~\ref{fig:Case1}c,d). We plot the error obtained with both schemes for the element Q8 in Figure \ref{fig:Case1Q8error}. It is evident that the integration point-based approach reproduces the analytical result more accurately. 

\begin{figure}[H]
        \centering
        \begin{subfigure}[h]{0.49\textwidth}
                \centering
                \includegraphics[scale=0.8]{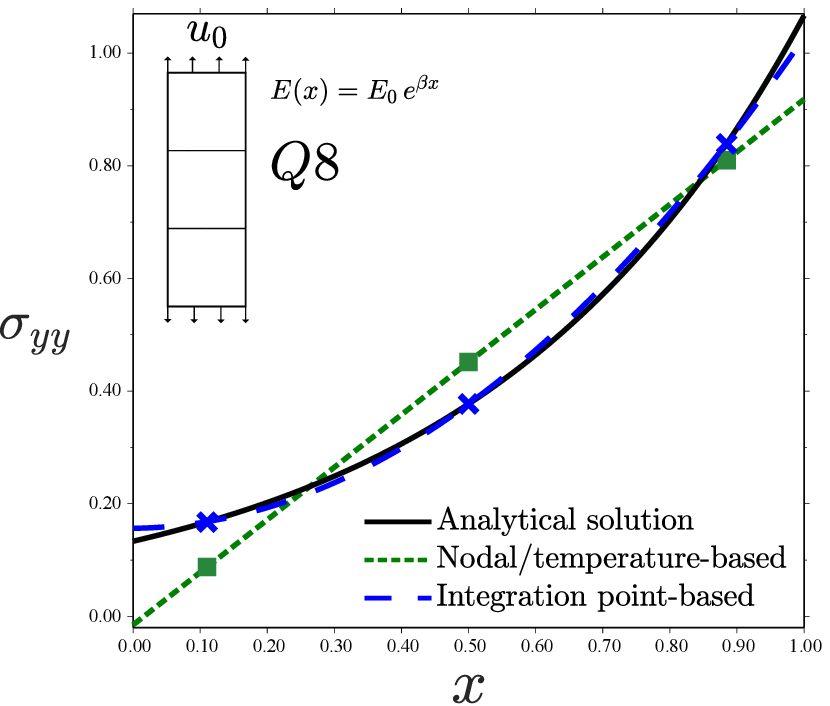}
                \caption{}
                \label{fig:Case1Q81elementA}
        \end{subfigure}
        \begin{subfigure}[h]{0.49\textwidth}
                \centering
                \includegraphics[scale=0.8]{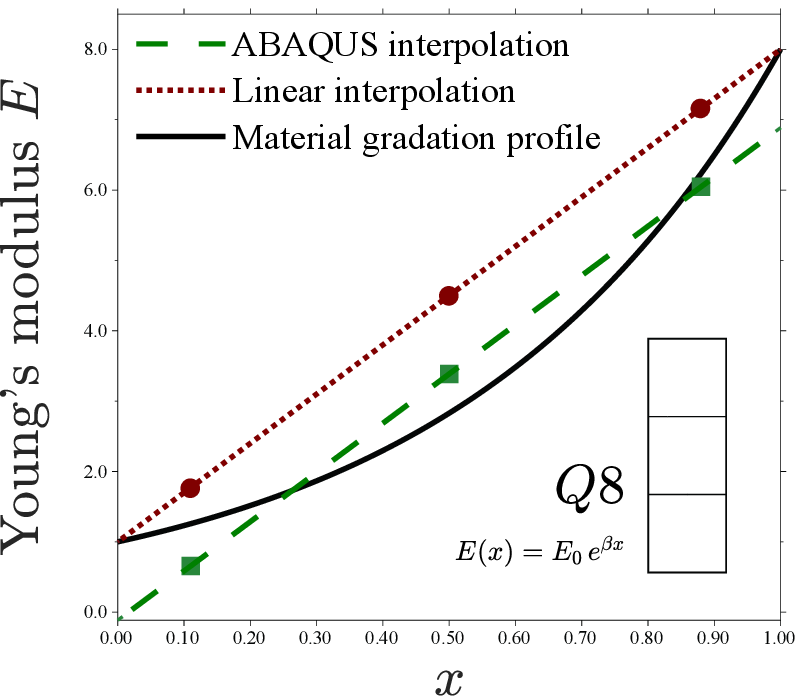}
                \caption{}
                \label{fig:Case1Q81elementB}
        \end{subfigure}\\
        \vspace{10pt}
        \begin{subfigure}[h]{0.5\textwidth}
                \centering
                \includegraphics[scale=0.8]{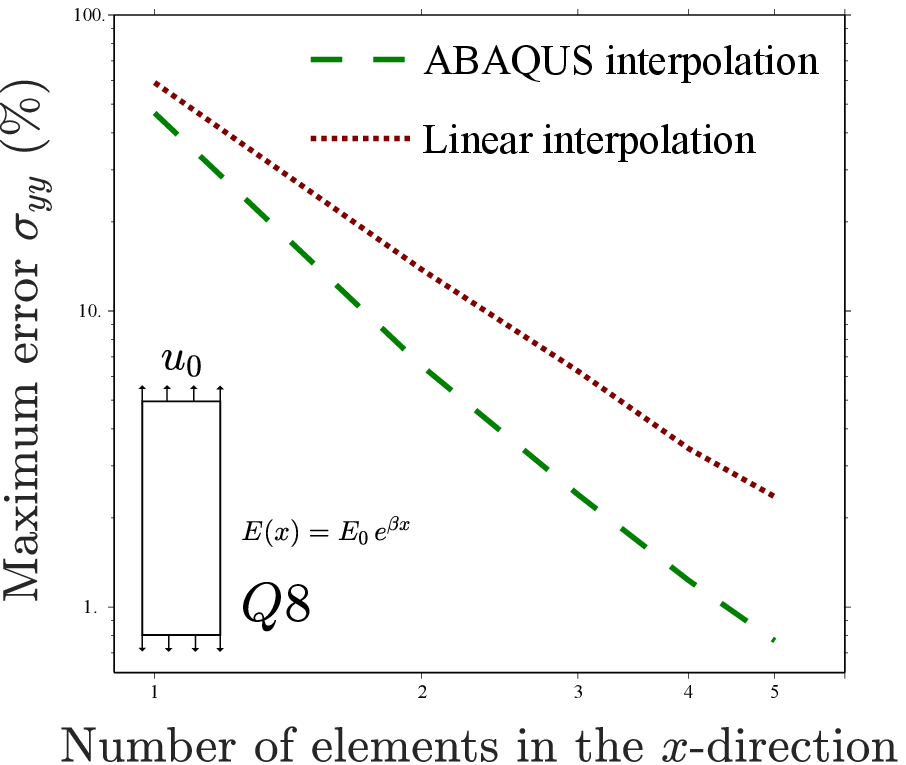}
                \caption{}
                \label{fig:Case1Q81elementC}
        \end{subfigure}
        \caption{Uniform displacement perpendicular to the material gradient, (\textbf{a}) tensile stress for one Q8 element in the $x$-direction; (\textbf{b}) Young's modulus interpolation through different schemes; and (\textbf{c}) mesh-sensitivity error analysis---consistent units.}\label{fig:Case1Q81element}
\end{figure}

Further insight is gained by reproducing the analysis with a single element in the $x$-direction---see Figure \ref{fig:Case1Q81element}. Inspection of Figure \ref{fig:Case1Q81element}a reveals substantial differences between Gauss points-based and nodal-based implementations. The former reproduces precisely the analytical stress distribution, with~the exact result being obtained at the integration points. Contrarily, the approximation is much poorer when the material gradient is implemented via the temperature. Differences are particularly significant at the left side of the specimen, where the error is on the order of 50\%. 

Remarkably, \emph{negative} stresses are predicted for $x=0$. These non-physical compressive stresses arise as a consequence of the particularities of ABAQUS' criterion for interpolating nodal temperatures, which does not correspond to the linear interpolation outlined in Figure \ref{fig:SketchLinearApprox}. In ABAQUS, the nodal temperature values are multiplied by certain weights, such that the temperature $T$ at an integration point $i$ is given by 
\begin{equation}
T_i = \sum_{j=1}^m T_j W_{ij} \,\,\,\,\, \text{with} \, i=1, \cdots, n.
\end{equation}

Here, $T_j$ is the temperature in node $j$, $W_{ij}$ the weight associated with the nodal temperature $j$ and integration point $i$, and $n$ and $m$ respectively denote the total number of nodes and integration points. The specific values of $W_{ij}$ depend on a number of numerical considerations and can be easily obtained by means of a one-element model. This criterion is motivated by numerical convergence in thermomechanical problems, as it smoothens localized temperature peaks. The resulting variation in the elastic properties within the element is shown in Figure \ref{fig:Case1Q81element}b. Differences with a direct linear interpolation are evident. The weighting procedure implemented in ABAQUS brings non-physical values of $E$, but it shows a better agreement with the material gradation profile at the Gauss integration points. In turn, this better approximation of $E(x)$ reduces the error in the computation of the stresses, as quantified in Figure \ref{fig:Case1Q81element}c as a function of the number of elements. The log--log plot of Figure \ref{fig:Case1Q81element}c shows that the weighted interpolation of ABAQUS exhibits a smaller maximum error in the computation of $\sigma_{yy}$ at the Gauss points, as well as a faster convergence rate. Consequently, the error intrinsic to a temperature-based graded element is magnified if a standard linear interpolation is used and, therefore, the conclusions of the present study are even more relevant to finite element codes that employ non-weighted interpolations of nodal temperatures. Recall that the integration point-based scheme presented in Section \ref{Sec:GaussBased} captures the analytical solution at the Gauss points with a single element.

\subsection{Uniform Traction Perpendicular to the Material Gradient Direction}
\label{Sec:TensionPerpendicular}

Consider now the case where the remote load is prescribed as a traction perpendicular to the elastic gradient---see Figure \ref{fig:BoundaryValueProblems}c. The Dirichlet boundary conditions of the problem read
\begin{equation}\label{Eq:BC1}
u_x (0,0) =0,
\end{equation}
\begin{equation}\label{Eq:BC2}
u_y (x,0) =0.
\end{equation}

In the case of a plate with infinite height, the only non-zero component of the Cauchy stress tensor is $\sigma_{yy}$. Following Refs. \cite{Erdogan1997,Kim2002}, a membrane resultant $N$ along the $x=w/2$ line can be defined as a function of the remote stress $\sigma_0$ and the width,
\begin{equation}
N= \sigma_0 w.
\end{equation}

The compatibility condition $\partial^2 \varepsilon_{yy} / \partial x^2 =0$ requires the strain component to be of the form
\begin{equation}
\varepsilon_{yy} (x) = Ax+B,
\end{equation}
and, consequently, one can readily obtain the stress field by considering the exponential elastic modulus variation assumed (\ref{eq:Evariation}) and making use of Hooke's law as
\begin{equation}
\sigma_{yy}(x)= E_0 e^{\beta x} (Ax+B).
\end{equation}

The coefficients $A$ and $B$ are obtained by solving
\begin{equation}
\int_0^w \sigma_{yy} (x) \text{d}x = N,
\end{equation}
\begin{equation}
\int_0^w \sigma_{yy} (x) x \text{d}x = N\frac{w}{2},
\end{equation}

\noindent such that
\begin{equation}
A=\frac{\beta N}{2E_0}\left(\frac{w\beta^2e^{\beta w}-2\beta e^{\beta w}+w\beta^2+2\beta}{e^{\beta w}\beta^2 w^2-e^{2\beta w}+2e^{\beta w}-1}\right),
\end{equation}
\begin{equation}
B=\frac{\beta N}{2E_0} \cdot \frac{e^{\beta w}[e^{\beta w}(-w^2\beta^2+3\beta w-4)+w^2\beta^2-2\beta w+8]-\beta w-4}{(e^{\beta w}-1)(e^{\beta w}\beta^2w^2-e^{2\beta w}+2e^{\beta w}-1)}.
\end{equation}

The displacement solution can be readily obtained by making use of the strain-displacement relations and applying the boundary conditions (\ref{Eq:BC1})--(\ref{Eq:BC2}),
\begin{equation}
u_x (x,y)=\nu \left(\frac{A}{2} x^2 + Bx \right) - \frac{A}{2} y^2,
\end{equation}
\begin{equation}
u_y (x,y)=\left( Ax + B \right) y.
\end{equation}

\begin{figure}[H]
\makebox[\linewidth][c]{%
        \begin{subfigure}[b]{0.45\textwidth}
                \centering
                \includegraphics[scale=0.8]{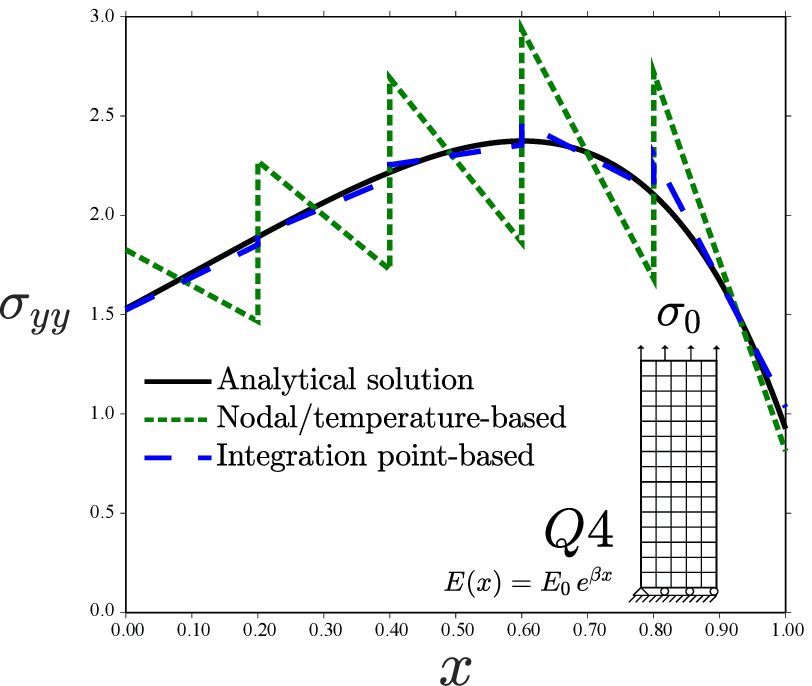}
                \caption{}
                \label{fig:Case2Q4}
        \end{subfigure}
        \begin{subfigure}[b]{0.45\textwidth}
                \raggedleft
                \includegraphics[scale=0.8]{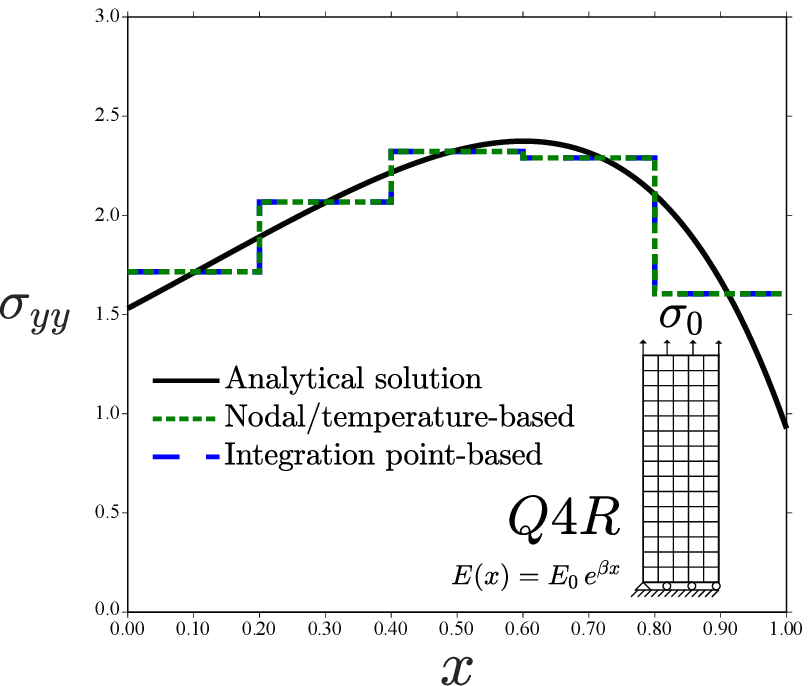}
                \caption{}
                \label{fig:Case2Q4R}
        \end{subfigure}}\\

\makebox[\linewidth][c]{%       
        \begin{subfigure}[b]{0.45\textwidth}
                \centering
                \includegraphics[scale=0.8]{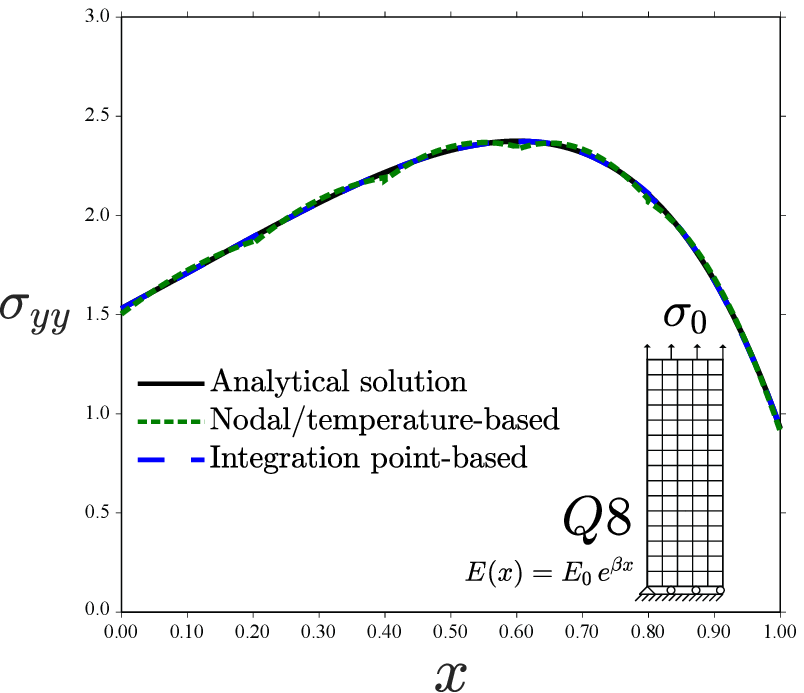}
                \caption{}
                \label{fig:Case2Q8}
        \end{subfigure}
        \begin{subfigure}[b]{0.45\textwidth}
                \raggedleft
                \includegraphics[scale=0.8]{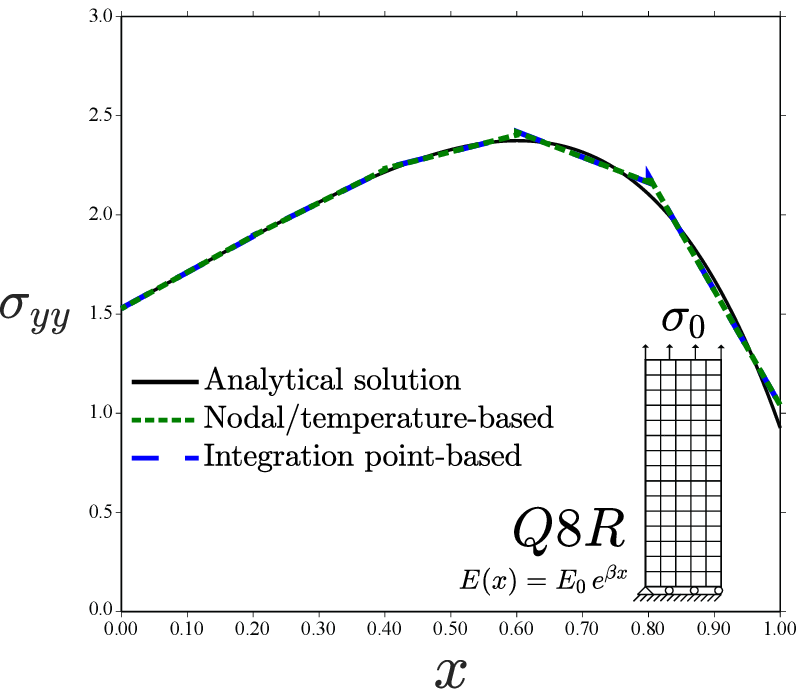}
                \caption{}
                \label{fig:Case2Q8R}
        \end{subfigure}
        }       
        \caption{Uniform traction perpendicular to the material gradient for different kinds of elements: (\textbf{a}) Q4; (\textbf{b}) Q4R; (\textbf{c}) Q8; and (\textbf{d}) Q8R---consistent units.}\label{fig:Case2}
\end{figure}

The analytical solution in the middle line $y=h/2$ is compared with the numerical predictions for $\sigma_0=2$. Results are shown in Figure \ref{fig:Case2} for a uniform mesh of 48 plane stress elements.

\begin{figure}[H]
\centering
\includegraphics[scale=1.4]{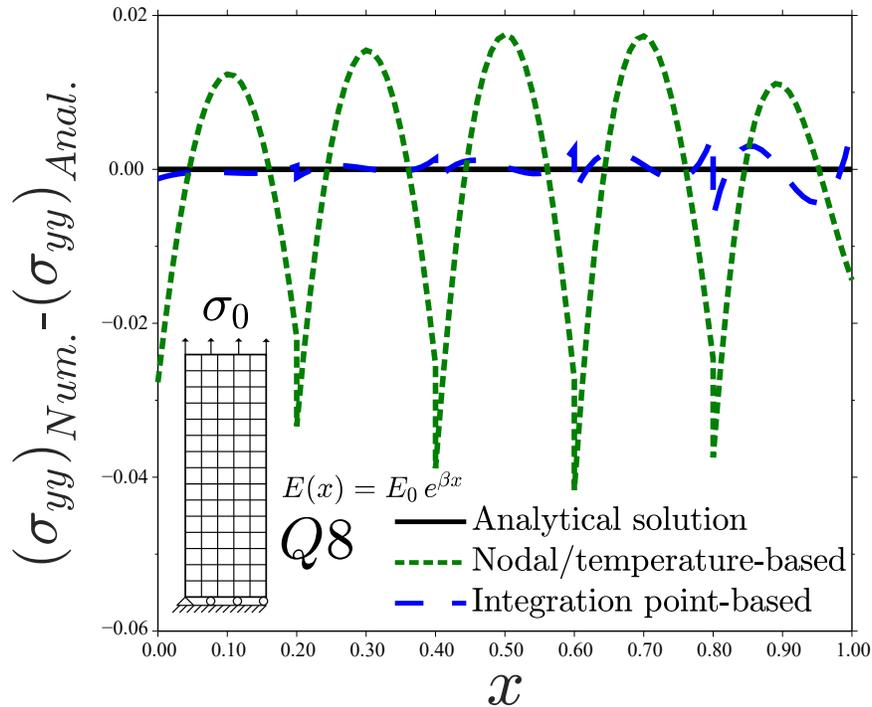}
\caption{Uniform traction perpendicular to the material gradient. Error analysis for the Q8 element---consistent units.}
\label{fig:Case2Q8error}
\end{figure}

Differences between the integration point-based scheme and the nodal/temperature-based implementation are particularly significant for the case of linear elements with full integration (Q4, Figure \ref{fig:Case2}a). Sampling the material gradient directly at the Gauss points leads to a good agreement with the analytical solution. However, using temperature-based properties renders the homogeneous element solution. Both the analytical and Gauss point-based solutions show an increasing $\sigma_{yy}$ along $x$ within those elements close to the left edge. Contrarily, the inverse response is observed when using temperature-dependent properties, as the strain field decreases with $x$ and $E$ is constant element-wise. The Q4 element predicts in all cases a linear variation of $\sigma_{yy}$ within each element for the quadratic displacement solution under consideration. On the other hand, identical predictions between graded element schemes are obtained when using linear elements with reduced integration (Q4R, Figure \ref{fig:Case2}b). This is unlike the case of a prescribed displacement (see~Figure \ref{fig:Case1}b), as the error in the approximation of the strain field $\varepsilon_{yy}$ (exact at the integration points only for the Gauss points-based scheme) is compensated. Quadratic elements (Q8 and Q8R) introduce an element-wise variation of $E$ in both approaches and, therefore, differences appear to be smaller than in linear elements---see~Figure~\ref{fig:Case2}c,d. The~error in the approximation is shown in Figure \ref{fig:Case2Q8error} for the Q8 element case. When using full integration, the Gauss point-based approach outperforms the temperature-based, generalized isoparameteric graded element.

\begin{figure}[H]
        \centering
        \begin{subfigure}[h]{0.49\textwidth}
                \centering
                \includegraphics[scale=0.8]{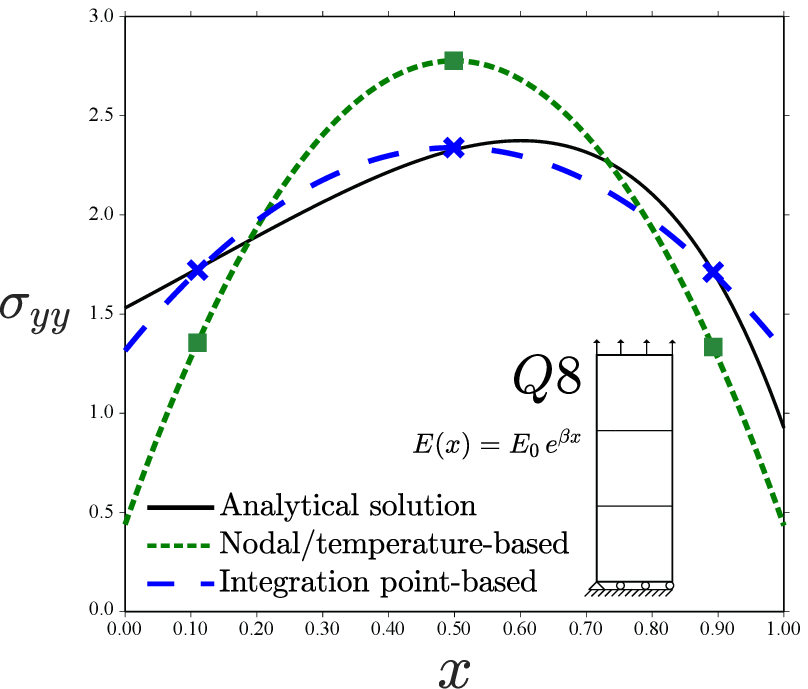}
                \caption{}
                \label{fig:Case2Q81elementA}
        \end{subfigure}
        \begin{subfigure}[h]{0.49\textwidth}
                \centering
                \includegraphics[scale=0.8]{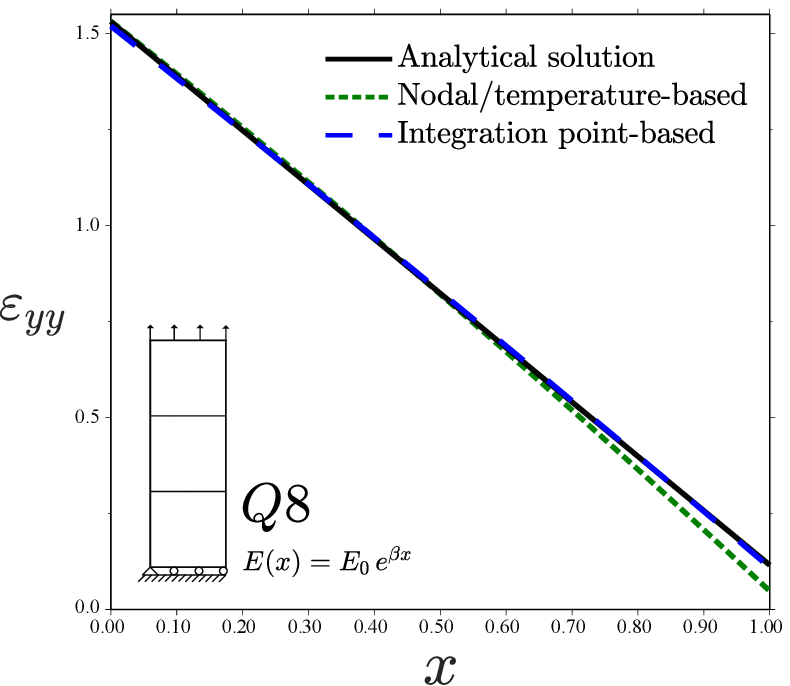}
                \caption{}
                \label{fig:Case2Q81elementB}
        \end{subfigure}
        \caption{Uniform traction perpendicular to the material gradient, tensile (\textbf{a}) stress and (\textbf{b}) strain for one Q8 element in the $x$-direction---consistent units.}\label{fig:Case2Q81element}
\end{figure}

Further insight is gained by analysing a very coarse mesh with a single element in the $x$-direction; results are shown in Figure \ref{fig:Case2Q81element}. Regarding the stress (Figure \ref{fig:Case2Q81element}a), the prediction obtained from the integration point-based implementation of graded elements agrees well with the analytical solution, being exact at the Gauss points (symbols). On the other hand, the approximation via a nodal/temperature-based approach introduces a significant source of error (larger than 20\% in all the integration points). Moreover, when the load is prescribed as a traction, the approximation of the material gradient also influences the strain field, see Equations (\ref{Eq:ElSystem})--(\ref{Eq:Ke}). As shown in Figure \ref{fig:Case2Q81element}b, a~better approximation is attained if the material properties are sampled directly at the Gauss points. Both schemes differ at the edges with the analytical solution for a plate of infinite height.

\subsection{Uniform Traction Parallel to the Material Gradient Direction}
\label{Sec:TensionParallel}

\begin{figure}[H]
        \centering
        \begin{subfigure}[h]{0.49\textwidth}
                \centering
                \includegraphics[scale=0.85]{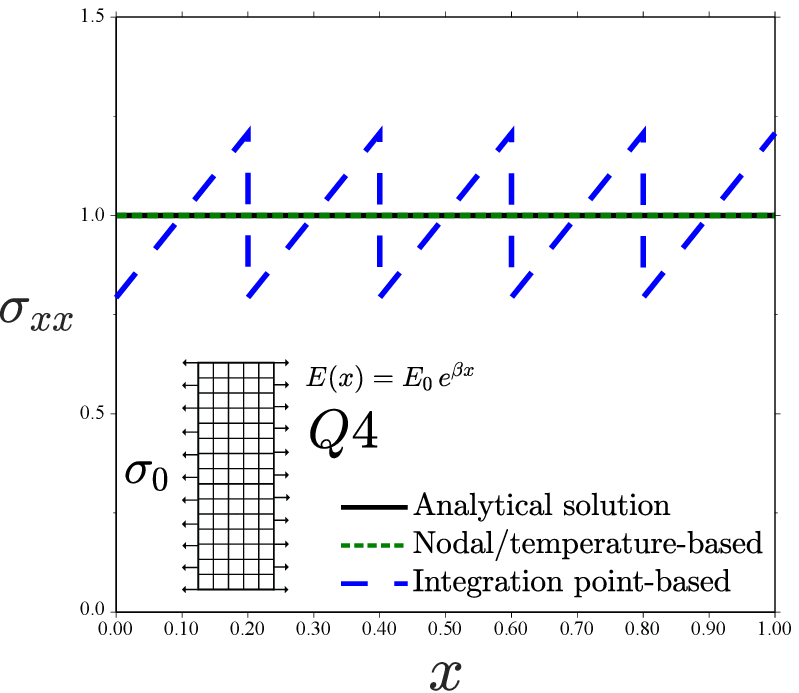}
                \caption{}
                \label{fig:Case3Q4}
        \end{subfigure}
        \begin{subfigure}[h]{0.49\textwidth}
                \centering
                \includegraphics[scale=0.85]{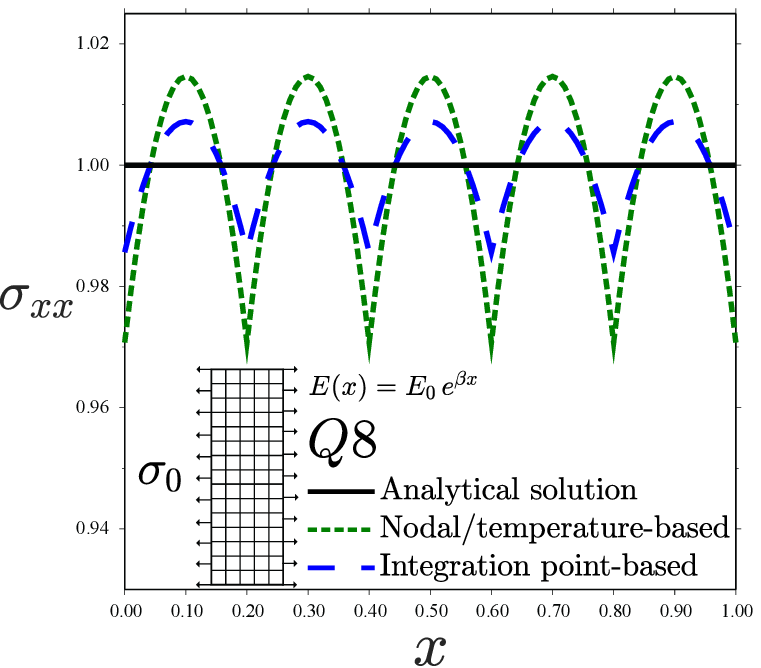}
                \caption{}
                \label{fig:Case3Q8}
        \end{subfigure}
        \caption{Uniform traction parallel to the material gradient for different kinds of elements: (\textbf{a}) Q4 and (\textbf{b}) Q8---consistent units.}\label{fig:Case3}
\end{figure}

The last case study involves a functionally graded plate subjected to traction in the $x$-direction, parallel to the elastic gradient, see Figure \ref{fig:BoundaryValueProblems}d. Under those conditions, the normal stress component equals the applied stress
\begin{equation}
\sigma_{xx} (x,y) = \sigma_0
\end{equation}

\noindent if Poisson's ratio is made equal to zero, $\nu=0$. The results obtained for a uniform mesh of 75 plane stress elements are shown in Figure \ref{fig:Case3}. 

Contrarily to what has been observed so far, the Q4 results (Figure \ref{fig:Case3}a) show that the nodal/temperature-based approach outperforms the integration point-based counterpart. Hooke's law requires the strains to vary according to an inverse exponential distribution to obtain a constant stress for an exponentially varying $E$. However, linear elements predict a constant strain field and, consequently, an effectively homogeneous element will predict a constant stress. This trend is inverted for the case of a quadratic element with full integration (Q8). As shown in Figure \ref{fig:Case3}b, again the use of a Gauss point-based graded element formulation approximates the analytical solution better. The results pertaining to reduced integration elements (Q4R and Q8R), not shown for brevity, reveal a perfect agreement with the analytical solution in all cases. Thus, reduced integration improves precision in this specific case study as the resulting stress is constant---either because there is a single integration point (Q4R), leading to a constant $\varepsilon_{xx}$ and $E$, or because both $\varepsilon_{xx}$ and $E$ are element-wise linear (Q8R). 

\section{Conclusions}

We have explored the influence of element order, integration scheme and graded element formulation in the finite element analysis of functionally graded materials (FGMs). Two graded element formulations are presented to account for the variation in space of material properties: nodal and integration point based gradations. The nodal based variation is implemented by defining temperature-dependent properties with a zero thermal expansion coefficient, a simple approach that enables the use of this scheme in commercial finite element packages. Important insight is gained by solving, analytically and numerically, three boundary value problems involving remote tractions and displacements, applied parallel and perpendicular to the material gradation direction. 

Results reveal that integration point-based graded elements generally outperform a nodal-based implementation through temperature-dependent properties. The former approximates better the analytical solution in all the boundary value problems considered if quadratic shape functions are used. A much finer mesh is needed to attain a similar degree of precision with a nodal-based approach. However, the temperature-based generalized isoparametric graded element is more accurate when linear elements are employed and the traction is applied parallel to the direction of material gradation. These observations are inherent to the interpolation of nodal temperatures with shape functions that are one order lower than those employed for the displacement field, as done in most finite element codes to ensure an equivalent variation of thermal and mechanical strains. In addition, in the case of the commercial package ABAQUS, the nodal temperature averaging criterion employed can lead to non-physical results. 

The results presented have implications in the analysis of functionally graded structures, both in terms of computation time and local precision, particularly relevant for fracture studies. A user subroutine for ABAQUS is presented to overcome the number of drawbacks identified with the most popular graded finite element implementation. The user subroutine can be downloaded from www.empaneda.com/codes. 
 
\section{Acknowledgments}
\label{Sec:Acknowledgments}

The author gratefully acknowledges financial support from the People Programme (Marie Curie Actions) of the European Union's Seventh Framework Programme (FP7/2007-2013) under REA grant agreement No. 609405 (COFUNDPostdocDTU).




\bibliographystyle{elsarticle-num}
\bibliography{library}

\end{document}